\documentclass[reprint, amsmath, amssymb, prl]{revtex4-1}

\pdfoutput=1

\usepackage[pdftex]{graphicx}
\usepackage{amsmath}
\usepackage{amssymb}

\newcommand{\rhov}{\boldsymbol{\rho}}

\newcommand{\rv}{\textbf{r}}

\newcommand{\rpv}{\textbf{r}'}

\newcommand{\Ev}{\textbf{E}}
\newcommand{\fv}{\textbf{f}}
\newcommand{\Fv}{\textbf{F}}

\newcommand{\Jv}{\textbf{J}}

\newcommand{\Vv}{\textbf{V}}

\newcommand{\xv}{\textbf{x}}

\newcommand{\Lambdav}{\boldsymbol{\Lambda}}
\newcommand{\zerov}{\textbf{0}}

\newcommand{\Gd}{\overline{\textbf{G}}\left(\rv,\rpv\right)}
\newcommand{\Gdkap}{\overline{\textbf{G}}_\kappa\left(\rv,\rpv\right)}

\newcommand{\Id}{\overline{\textbf{I}}}

\newcommand{\Dd}{\overline{\textbf{D}}}
\newcommand{\Pd}{\overline{\textbf{P}}}

\newcommand{\Vd}{\overline{\textbf{V}}}
\newcommand{\Md}{\overline{\textbf{M}}}
\newcommand{\Sd}{\overline{\textbf{S}}}

\newcommand{\Zd}{\overline{\textbf{Z}}}
\newcommand{\zerod}{\overline{\textbf{0}}}

\begin{document}
\title{Casimir Force for Arbitrary Objects Using the Argument Principle and Boundary Element Methods}
\thanks{This work was supported in part by the Research Grants Council of Hong Kong (GRF 711609, 711508, and 711511), in part by the University Grants Council of Hong Kong (Contract No. AoE/P-04/08), in part by the University of Illinois at Urbana-Champaign.}
\author{P. R. Atkins}
\affiliation{Center for Computational Electromagnetics and
Electromagnetics Laboratory, Department of Electrical and Computer Engineering,
University of Illinois at Urbana-Champaign, Urbana, IL 61801 USA.}
\author{W. C. Chew}
\email{w-chew@illinois.edu}
\affiliation{Center for Computational Electromagnetics and
Electromagnetics Laboratory, Department of Electrical and Computer Engineering,
University of Illinois at Urbana-Champaign, Urbana, IL 61801 USA.}
\author{Q. I. Dai}
\affiliation{The University of Hong Kong, Pok Fu Lam Road, Hong Kong.}
\author{W. E. I. Sha}
\affiliation{The University of Hong Kong, Pok Fu Lam Road, Hong Kong.}


%


\begin{abstract}
Recent progress in the simulation of Casimir forces between various objects has allowed traditional computational electromagnetic solvers to be used to find Casimir forces in arbitrary three-dimensional objects.  The underlying theory to these approaches requires knowledge and manipulation of quantum field theory and statistical physics.  We present a calculation of the Casimir force using the method of moments via the argument principle.  This simplified derivation allows greater freedom in the moment matrix where the argument principle can be used to calculate Casimir forces for arbitrary geometries and materials with the use of various computational electromagnetic techniques.
\end{abstract}

\maketitle

\section{Introduction}
The ability to model the bulk effect of molecular forces has become increasingly important with the development of micro- and nanoelectromechanical systems (MEMS and NEMS).  The static and dynamic behaviors of such devices involve the evaluation of molecular and electrodynamic forces.  The Casimir force can simulate them more accurately than a simpler form of the van der Waals force.  As such, a simple and robust method of calculating the Casimir force is desirable.  Recent methods have been derived based around boundary element methods\cite{PhDThesis:Reid,PhDThesis:Reid2}.  The Reid, Rodriguez, White, and Johnson (RRWJ) method allows the use of computational electromagnetic (CEM) solvers using the method of moments to find the Casimir force for arbitrary three-dimensional objects of arbitrary medium.  This presents an attractive method for calculating the Casimir force.  However, this method uses a traditional method of moments matrix which suffers from low frequency breakdown and mesh density breakdown.  Instead of deriving through the path integral, we show a new method that uses the argument principle with boundary element methods to arrive at an improved formulation.

Calculating the Casimir force using the argument principle was originally applied by Van Kampen \emph{et al.}\cite{PhDThesis:VanKampen} for a much simpler problem.  With the new derivation, we can choose from a wide range of modern CEM formulations that overcome shortcomings with the minimum of work.  In doing so, we expand beyond the class of problems that can currently be solved using the RRWJ method.  A simple demonstration of this is done using a formulation that avoids the low frequency breakdown for perfect electrical conductors (PEC), the Augmented-Electric Field Integral Equation (A-EFIE)\cite{PhDThesis:Qian} approach.

\section{Derivation}
The RRWJ method starts out using a path-integral expression for the Casimir energy of arbitrary PEC objects at $T=0$\cite{PhDThesis:Reid, PhDThesis:Li}.
\begin{align}
\mathcal{E} = -\frac{\hbar c}{2\pi} \int_0^\infty d\kappa \frac{\mathcal{Z}(\kappa)}{\mathcal{Z}_\infty(\kappa)} \label{EGJKstart}
\end{align}
where
\begin{align}
\mathcal{Z}(\kappa) = \int \mathcal{D}\Jv(\rv) e^{-1/2 \int d\rv \int d\rpv \Jv(\rv)\cdot \Gdkap \cdot \Jv(\rpv)}
\end{align}
and
\begin{align}
\Gdkap = \left[ \Id + \frac{\nabla\nabla'}{\kappa^2}\right] \frac{e^{-\kappa \left| \rv-\rpv \right|}}{4\pi\left| \rv-\rpv \right|}
\end{align}
Here, the functional integration $\mathcal{Z}$ is performed over all possible configurations of the current $\Jv(\rv)$ on the surfaces of the objects and $\Gdkap$ is the dyadic Green's function with the Wick rotated imaginary frequency.  $\mathcal{Z}_\infty$ is computed similarly as $\mathcal{Z}$ except with all objects removed to infinite separation.  By expanding the continuous current distribution $\Jv(\rv)$ over a finite discrete set of basis functions $\fv_{m}$, defined over the surfaces of the objects, the RRWJ method reduces to
\begin{align}
\mathcal{E} = \frac{\hbar c}{2\pi}\int_0^\infty d\kappa \ln \frac{\det \Md(\kappa)}{\det \Md_\infty(\kappa)} \label{EGJKfinal}
\end{align}
The elements of the matrix $\Md$, being
\begin{align}
\left[ \Md \right]_{ij} = \int \int \fv_i(\rv)\cdot \Gdkap \cdot \fv_j(\rpv) d\rpv d\rv
\end{align}
are of the same form as the matrix elements of the traditional CEM impedance matrix for the method of moments.  Equation \eqref{EGJKfinal} is derived from Equation \eqref{EGJKstart} assuming a specific action for the path integral and properties of the matrix $\Md$.  Attempts to modify the resulting impedance matrix used in Equation \eqref{EGJKfinal} requires a new path integral formulation as well as the assumption that the matrix $\Md$ be a positive or negative definite matrix.  In addition to PEC objects, this formulation can handle homogeneous dielectric objects\cite{PhDThesis:Reid2}.

However, greater freedom is needed to have a matrix $\Md$ that handles a wider class of problems, like inhomogeneous dielectric objects, or addresses numerical instabilities like low frequency integration points.  To ensure low frequency stability, the loop-tree decomposition\cite{PhDThesis:ChewBook2} can be used, whereby the impedance matrix is the result of a similarity transform of the traditional method of moments impedance matrix.  This transform, achieved using left and right matrix multiplications, is canceled out when the determinant is normalized with $\Md_\infty$, thus resulting in the mathematical equivalence of Equation \eqref{EGJKfinal}.  Other low frequency stable methods like A-EFIE, unfortunately, are not symmetric positive definite nor represent a similarity transform of the predefined matrix $\Md$.  Instead of evaluating a new path integral, we approach the problem using the argument principle.

Note that the Casimir force results from the relative perturbation of the quantum vacuum fields by the pertinent objects.  The objects force the fluctuating vacuum fields to conform to the appropriate boundary conditions and, in doing so, change the energy density of the vacuum.  If one determines the eigenfrequencies of the field configurations that satisfy the geometry's boundary conditions, then the unnormalized energy of the geometry can be found via\cite{PhDThesis:Barash, PhDThesis:Milonni}
\begin{align}
\mathcal{E} = \sum_\omega \frac{1}{2} \hbar \omega \label{CasEnerg}
\end{align}
Previous derivations of the Casimir force have made use of the argument principle to find the eigenmodes of the vacuum fields to define the energy density\cite{PhDThesis:VanKampen, PhDThesis:Barash, PhDThesis:Langbein, PhDThesis:Schram, PhDThesis:Lambrecht}.  To calculate the Casimir energy from these dispersion equations, we first note that the argument principle states that
\begin{align}
\frac{1}{2\pi i} \oint \phi(\omega) \frac{d}{d\omega} \ln f(\omega) d\omega = \sum_i{\phi(\omega_{0,i})} - \sum_j{\phi(\omega_{\infty,j})}
\end{align}
where $\omega_{0,i}$ are the zeros and $\omega_{\infty,j}$ are the poles of the function $f(\omega)$ inside the contour of integration.  Noting that the Casimir energy of a geometry is given by Equation \eqref{CasEnerg}, we can relate the Casimir energy with the contour integral from above.  Using integration by parts, the Wick rotation where $\omega = ic\kappa$, and folding the integral, we find
\begin{align}
\mathcal{E} = \frac{\hbar c}{2\pi} \int_0^\infty d\kappa \ln f(ic\kappa) \label{CasEnergIntegral}
\end{align}
where the zeros of $f(\omega)$ correspond to the eigenmodes of the structure.  It is necessary, in calculating the Casimir energy, to renormalize the vacuum energy by subtracting off an appropriate normalization energy represented by modes $\omega_{norm}$.  That is,
\begin{align}
\mathcal{E} = \sum_{\omega,\omega_{norm}} \frac{1}{2} \hbar \left[ \omega - \omega_{norm} \right]
\end{align}
Following the same process, the Casimir energy is represented as
\begin{align}
\mathcal{E} = \frac{\hbar c}{2\pi} \int_0^\infty d\kappa \ln \frac{f(ic\kappa)}{f_{norm}(ic\kappa)} \label{CasEnergIntegralNorm}
\end{align}
The theoretical problem is then to find some function $f(\omega)$ and its associated normalization $f_{norm}(\omega)$, taken as the geometry when the objects are infinitely separated, that are zero at the eigenfrequencies of the Casimir problem.  This has been previously done by finding the dispersion relationship of the geometry using primarily closed-form analysis.  This could only be applied to a small number of geometries where exact dispersion relations could be derived like periodic gratings or dielectric slabs and was not extended to arbitrary geometries.

The preceeding derivation relies on the physical interpretation of the natural modes of the problem and the summation over their energies to arrive at the Casimir energy in Equation \eqref{CasEnerg}.  As such, the above is valid for a lossless system where the eigenfrequencies are real, but the physical meaning of these eigenfrequencies becomes indeterminate in a lossy system with complex eigenfrequencies.  Despite this, the use of the argument principle using lossy materials still seemingly resulted in the correct Casimir force\cite{PhDThesis:Milonni, PhDThesis:Ginzburg, PhDThesis:Lamoreaux2}.  Only recently have papers have been published that rigorously explain why\cite{PhDThesis:Milonni2, PhDThesis:Milonni3, PhDThesis:Sernelius}.

The traditional CEM formulation for PEC objects is derived assuming that the fields arising from currents induced on the surface of a PEC must cancel any excitation fields that are present.  This boundary condition leads to the Electric Field Integral Equation (EFIE),
\begin{align}
-\hat{n}\times\Ev^s(\rv) = \hat{n}\times\Ev^i(\rv) = \hat{n} \times i\omega\mu \int d\rpv \Gd \cdot \Jv(\rpv) \label{EFIE}
\end{align}
where $\Gd$ is the non-Wick rotated dyadic Green's function, $\Ev^s$ is the scattered electric field and $\Ev^i$ is the incident exciting electric field.  The method of moments is used to derive an impedance matrix to relate surface currents to the excitation fields.  That is,
\begin{align}
\Zd \cdot \Jv = \Vv \label{MoMEq}
\end{align}
where $\Zd$ is the matrix representation of the real frequency Green's operator and $\Jv$ and $\Vv$ are the vector representations of the current $\Jv(\rv)$ and exciting fields respectively.  As previously stated, the eigenmodes of the quantum vacuum are those where the fluctuating fields satisfy the boundary conditions of the geometry.  In other words, eigenmodes exist where we have currents on the surface of the objects without any excitation fields.  These natural modes automatically satisfy the geometry's boundary conditions, and they also satisfy the relation,
\begin{align}
\Zd \cdot \Jv = \zerov
\end{align}
Being a singular matrix, we conclude that $f = \det \Zd = 0$ for the natural eigenfrequencies where the vacuum fields satisfy the boundary conditions without an external source.  Artificially enclosing the objects in a PEC cavity ensures that the eigenmodes of the system will be lossless and discrete.  By using an appropriate cavity dyadic Green's function in $\Zd$, $\det \Zd$ can be used as the dispersion relation in Equation \eqref{CasEnergIntegral} to find the energy of the bounded cavity.  Expanding the volume of the cavity to infinity and applying the normalization where the objects are infinitely separated, we reach Equation \eqref{CasEnergIntegralNorm}.

However, by using the cavity Green's function, the result is still incorrect and thus an infinitesimal loss, to enforce the Sommerfeld radiation condition\cite{PhdThesis:Chew}, is added to remove the reflections off the cavity walls at infinity.  In doing so, the resulting Green's function becomes the free-space Green's function, as it was in the original EFIE, and the RRWJ formulation is rederived.  Due to the infinitesimal loss, the eigenfrequencies of the EFIE are complex frequencies requiring one to justify its derivation for lossy systems.  The strength of this derivation is its large degree of flexibility.  As long as one finds a relationship that similarly relates the geometry's boundary conditions with the fields excited by induced and external sources, then the same procedure can be applied.  This allows us to rederive the RRWJ formulations for PEC and dielectric homogeneous objects in addition to more advanced CEM formulations that would otherwise require novel evaluations of the original path integral.

\begin{figure}[t]
\centering
\includegraphics[width=3.375in]{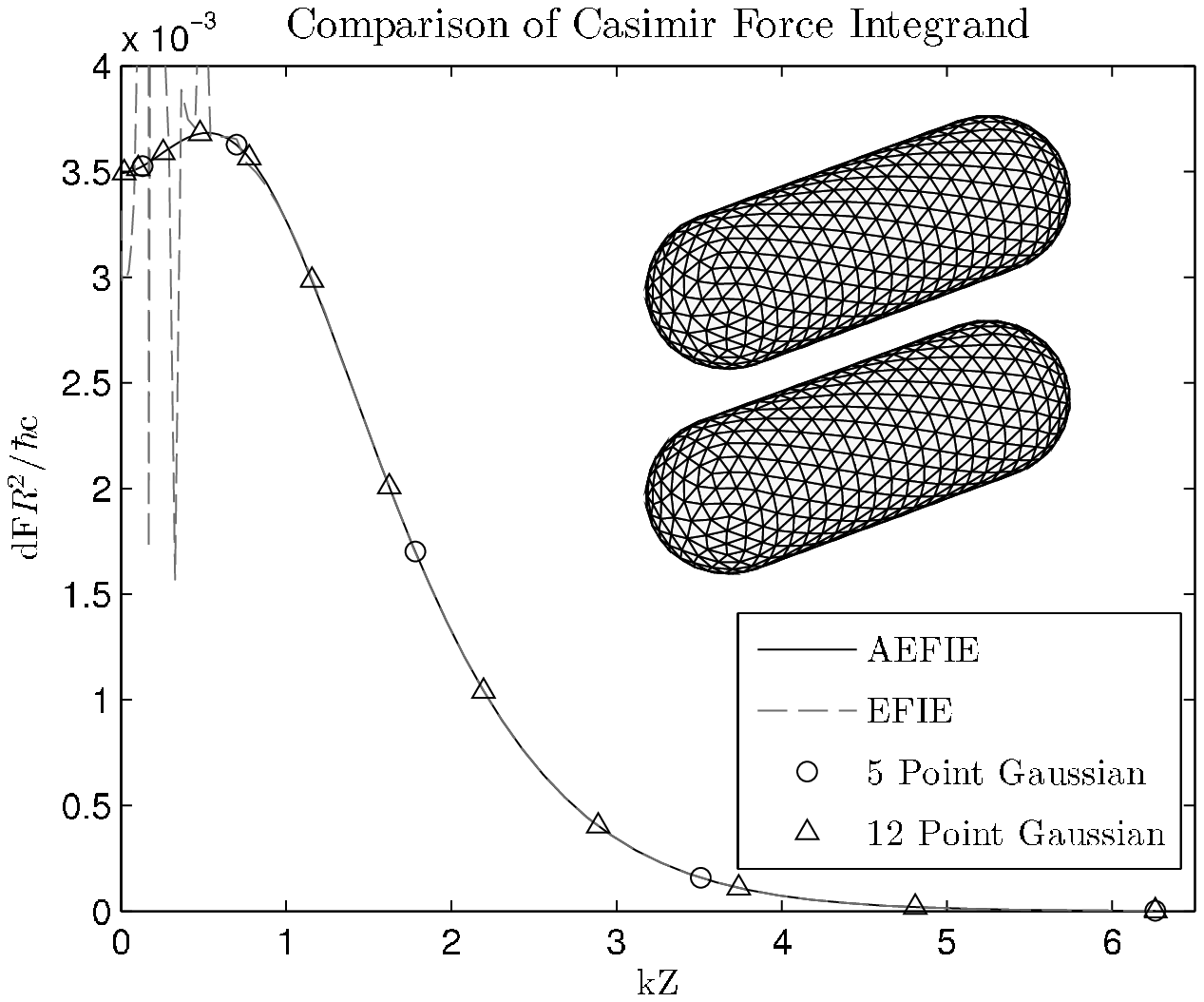}
\caption{Comparison of the integrand for the Casimir force integral between capsules for the EFIE and A-EFIE formulations using single precision.}
\label{ForceComp}
\end{figure}

One difficulty that arises with the numeric integration of the Casimir energy and force over the wavenumbers $\kappa$ is the low frequency breakdown.  Looking at the force spectrum for two PEC rounded cylinders, as shown in Figure \ref{ForceComp}, the spectrum is concentrated in the lower frequencies as is expected due to the normalization.  Evaluating the eigenvalues for the impedance matrices at lower frequencies becomes inaccurate due to the low frequency breakdown of the EFIE impedance matrix.  While increasing the precision of the calculations helps address this issue, it is still a problem depending upon the CEM formulation and class of problems being solved.  A better remedy is to use a more appropriate formulation for the impedance matrix, like A-EFIE, that will not suffer from the same breakdown.  A-EFIE starts from the normal EFIE formulation whose matrix representation in Equation \eqref{MoMEq} can be equivalently expressed as
\begin{align}
\left( ik_0\eta_0\Vd + \frac{\eta_0}{ik_0} \Sd \right) \cdot \Jv = \Vv
\end{align}
Assuming that the subspace is spanned by the Rao-Wilton-Glisson (RWG) basis\cite{PhDThesis:RWG}, the problem has $e$ RWG edges and $p$ patches.  The vector and scalar potential matrices above, $\Vd$ and $\Sd$ respectively, and an additional matrix representation of the scalar Green's function using patch basis are defined to be
\begin{align}
\Vd_{m,n} &= \mu_r \int_{S_m} \Lambdav_m(\rv) \cdot \int_{S_n} g(\rv,\rpv) \Lambdav_n(\rpv) dS'dS \\
\Sd_{m,n} &= \epsilon^{-1}_r \int_{S_m} \nabla \cdot \Lambdav_m(\rv) \cdot \int_{S_n} g(\rv,\rpv) \nabla' \cdot \Lambdav_n(\rpv) dS'dS \\
\Pd_{m,n} &= \epsilon^{-1}_r \int_{S_m} h_m(\rv) \int_{S_n} g(\rv,\rpv) h_n(\rpv) dS'dS
\end{align}
where $g(\rv,\rpv)$ is the homogeneous scalar Green's function, $\Lambdav(\rv)$ is an RWG basis that is not normalized against the edge length, and $h(\rv)$ is a pulse basis.  We can relate the patch matrix, $\Pd$, with the scalar potential matrix, $\Sd$, using the incidence matrix $\Dd$ by
\begin{align}
\Sd = \Dd^T\cdot\Pd\cdot\Dd
\end{align}
By accounting for current continuity,
\begin{align}
\Dd\cdot\Jv = ik_0c_0\rhov
\end{align}
we can derive the A-EFIE matrix system as
\begin{align}
\left[ \begin{matrix} \Vd & \Dd^T\cdot\Pd \\ \Dd & k_0^2\Id \end{matrix} \right] \cdot \left[ \begin{matrix} ik_0\Jv \\ c_0\rhov \end{matrix} \right] = \left[ \begin{matrix} \eta_0^{-1}\Vv \\ \zerov \end{matrix} \right] \label{AEFIE}
\end{align}

The given formulations for A-EFIE can be further improved upon and it is seen that the A-EFIE impedance matrix, $\Zd_A$, will have the function $f(\omega) = \det \Zd_A = 0$ at the eigenfrequencies of the geometry.  The A-EFIE impedance matrix can be used as a direct replacement for the EFIE impedance matrix in the Casimir calculations by use of the argument principle.  Thus, the Casimir energy and force become,
\begin{align}
\mathcal{E} &= \frac{\hbar c}{2\pi} \int_0^\infty d\kappa \ln \frac{\det \Zd_A(\kappa)}{\det \Zd_{A,\infty}(\kappa)} \label{CasEnergyA} \\
\Fv &= -\frac{\hbar c}{2\pi} \int_0^\infty d\kappa \nabla_i \ln \frac{\det \Zd_A(\kappa)}{\det \Zd_{A,\infty}(\kappa)} \label{CasForceA}
\end{align}
where $\nabla_i$ represents the derivatives with respect to a physical displacement of the $i$-th object (the one we wish to find the forces acting upon).  The integrand of the Casimir energy is found by solving for the eigenvalues of $\Zd_A$ and $\Zd_{A,\infty}$.
\begin{align}
\ln \left( \frac{\det \Zd_A(\kappa)}{\det \Zd_{A,\infty}(\kappa)} \right) = \sum_{n=1}^N{ \ln \left( \frac{\lambda_n}{\lambda_n^\infty} \right) }
\end{align}
where $\lambda_n$ and $\lambda_n^\infty$ are the eigenvalues of $\Zd_A$ and $\Zd_{A,\infty}$ respectively.  For the Casimir force, note that the integrand can be expressed as
\begin{align}
\nabla_i \ln \det \Zd_A(\kappa) = \sum_{n=1}^N{\alpha_n}
\end{align}
where $\alpha_n$ are the eigenvalues of the generalized eigenvalue problem
\begin{align}
\nabla_i \Zd_A \cdot \xv = \alpha \Zd_A \cdot \xv \label{GenEig}
\end{align}
It should be noted that the gradient acting upon $\Zd_A$ acts on all elements of the matrix, making the bottom block matrices zeros.  That is,
\begin{align}
\nabla_i \Zd_A &= \left[ \begin{matrix} \nabla_i \Vd & \Dd^T\cdot \nabla_i \Pd \\ \zerod & \zerod \end{matrix} \right]
\end{align}

\begin{figure}[t]
\centering
\includegraphics[width=3.375in]{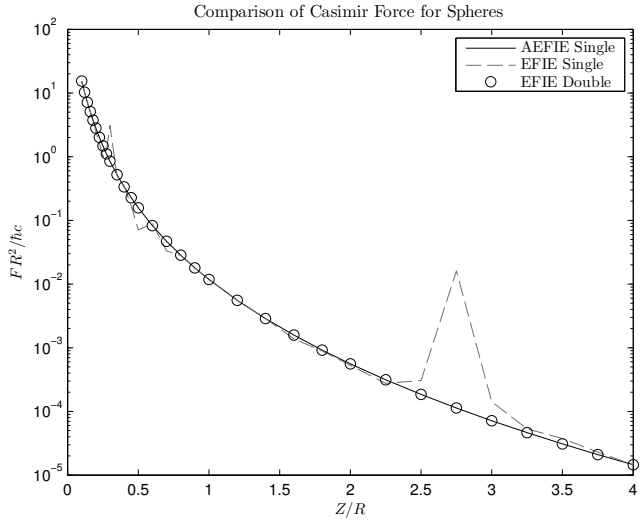}
\caption{Comparison of the Casimir force between two PEC spheres for the EFIE and A-EFIE formulations using single precision.}
\label{TotalForceComp}
\end{figure}

Using A-EFIE, we can more accurately calculate the low frequency energy spectrum than using EFIE while at the same time being able to solve the same set of problems as the original RRWJ formulation.  While in general, double precision can give satisfactory results, Figure \ref{ForceComp} shows that the A-EFIE and EFIE results start to differ as the frequency drops when using single precision for the force between two parallel PEC rounded cylinders of length to radius ratio of 6, and separation to radius ratio of 4.  The nominal dimensions are immaterial as the force and energy results are normalized with respect to the radius of the objects.  The result from the numerical integral is calculated using points indicated in the figure and the points with the largest contribution lie in the region of poor performance.  The deviation in the calculated integration is a result of the location of the Gaussian points but the error in the EFIE result can be greater than 100\% while A-EFIE remains in agreement.  Figure \ref{TotalForceComp} demonstrates how the total force between two PEC spheres can diverge over a range of values for the ratio between the separation and radius of the spheres.

\section{Conclusion}
This paper, using the argument principle, presents an alternative and novel derivation for the Casimir energy and force using boundary element methods.  As a result, a simple procedure can easily be adapted to different CEM methods to derive previous formulations or new ones.  For example, the characteristics of the Casimir force can present difficulties when using traditional EFIE.  The importance of the low frequency content of the integrand gives rise to erroneous results when relaxing the computational precision due to low frequency breakdown.  As a demonstration of the utility in using boundary element methods with the argument principle, for the first time, the low frequency A-EFIE was used to overcome the limitations of the EFIE formulation in Casimir force calculations.  This opens up possibilities for other improvements like using the mixed-form fast multipole algorithm with A-EFIE\cite{PhDThesis:Qian, PhDThesis:Jiang} or the use of other formulations like the PMCHWT method for homogeneous dielectrics\cite{PhDThesis:Poggio, PhDThesis:Chang, PhDThesis:Wu, PhDThesis:Medgyesi} (which can be used to rederive another result by RRWJ\cite{PhDThesis:Reid2}), the volume integral method for inhomogeneous dielectric objects\cite{PhDThesis:ChewSun}, and the layered medium Green's function\cite{PhDThesis:ChewMatrix} to help model systems with infinite slabs like a substrate.  Many of these formulations open a new class of geometries or scale of problems that would otherwise not be accessible using the boundary element methods found in current methods.

\bibliography{Atkins_Casimir}

\clearpage

\end{document}